\documentclass{article}
\usepackage{dcase2020,amsmath,graphicx,url,times,booktabs, tabularx}

\title{Guided multi-branch learning systems for sound event detection with sound separation}

 \name{Yuxin Huang$^{1,2}$,
       Liwei Lin$^{1,2}$,
       Shuo Ma$^{1,2}$,
       Xiangdong Wang$^{1*}$,
       Hong Liu$^{1}$,
       Yueliang Qian$^{1}$,
       }
 \secondlinename{	
       Min Liu$^{3}$,
       Kazushige Ouchi$^{3}$
       }
 \address{$^1$ Beijing Key Laboratory of Mobile Computing and Pervasive Device,\\
         Institute of Computing Technology, Chinese Academy of Sciences, Beijing, China,\\
         \{huangyuxin18g, linliwei17g\}@ict.ac.cn, ms\_hebut@163.com, \{xdwang, hliu, ylqian\}@ict.ac.cn\\
         $^2$ University of Chinese Academy of Sciences, Beijing, China\\
         $^3$ Toshiba China R\&D Center, Beijing, China,\\
         liumin@toshiba.com.cn, kazushige.ouchi@toshiba.co.jp\\
         *Corresponding author
  }

\begin{document}

\ninept
\maketitle

\begin{sloppy}

\begin{abstract}
In this paper, we describe in detail our systems for DCASE 2020 Task 4. The systems are based on the 1st-place system of DCASE 2019 Task 4, which adopts weakly-supervised framework with an attention-based embedding-level pooling module and a semi-supervised learning approach named guided learning. This year, we incorporate multi-branch learning (MBL) into the original system to further improve its performance. MBL uses different branches with different pooling strategies (including instance-level and embedding-level strategies) and different pooling modules (including attention pooling, global max pooling or global average pooling modules), which share the same feature encoder of the model. Therefore, multiple branches pursuing different purposes and focusing on different characteristics of the data can help the feature encoder model the feature space better and avoid over-fitting. To better exploit the strongly-labeled synthetic data, inspired by multi-task learning, we also employ a sound event detection branch. To combine sound separation (SS) with sound event detection (SED), we fuse the results of SED systems with SS-SED systems which are trained using separated sound output by an SS system. The experimental results prove that MBL can improve the model performance and using SS has great potential to improve the performance of SED ensemble system.
\end{abstract}


\section{Introduction}
\label{sec:intro}

DCASE 2020 task 4 \cite{dcase2020task4web} is the  follow-up to DCASE 2019 task 4 \cite{dcase2019task4web}. While DCASE 2019 task 4 targets on exploring the usage of weakly labeled data, unlabeled data and synthetic data in sound event detection (SED), DCASE 2020 task 4 encourages participants to combine sound separation with SED in addition to the same task in DCASE 2019. There are three subtasks in DCASE 2020 task 4: SED without sound separation, SED with sound separation and sound separation (using the SED baseline system). We participated in the first two subtasks. However, for the second subtask, we just use the baseline system for sound separation provided by the challenge organizer and focus on combination of sound separation and SED.

In this paper, we describe in detail our systems for the two subtasks we participated in DCASE2020 task 4. The systems are based on the first-place system of DCASE 2019 task 4 developed by Institute of Computing Technology (ICT), Chinese Academy of Sciences \cite{DCASE2019ICT}, which adopts the multiple instance learning framework with embedding-level attention pooling \cite{SDS} and a semi-supervised learning approach called guided learning \cite{Guidedlearning}. The multi-branch learning approach (MBL) \cite{MBL} is then incorporated into the system to further improve the performance. Multiple branches with different pooling strategies (embedding-level or instance-level) and different pooling modules (attention pooling, global max pooling or global average pooling) are used and share the same feature encoder. To better exploit the synthetic data with strong labels, inspired by multi-task learning \cite{Multitask}, a sound event detection branch is also added. Therefore, multiple branches pursuing different purposes and focusing on different characteristics of the data can help the feature encoder model the feature space better and avoid over-fitting. To incorporate sound separation into SED, we train models using output of the baseline system of sound separation and fuse the event detection results of models with or without sound separation.

\section{The DCASE 2019 task 4 system by ICT}
\label{sec:format}

Our systems for DCASE 2020 task 4 follows the framework of the DCASE 2019 task 4 system by ICT \cite{DCASE2019ICT}, which won the 1st place and the reproducible system award in the DCASE 2019 task 4 challenge. The system utilizes convolutional neural network (CNN) with embedding-level attention pooling module for weakly-supervised SED and uses disentangled features to solve the problem of unbalanced data with co-occurrences of sound events \cite{SDS}. To better use the unlabeled data jointly with weakly-labeled data, the system adopts a semi-supervised learning method named guided learning \cite{Guidedlearning}, which uses different models for the teacher model and student model to achieve different purposes implied in weakly-supervised SED. For the synthetic data, the system regards them as weakly annotated training data and the time stamps of sound events in the strong labels are not used. The system is trained by the DCASE 2019 training data, including weakly-labeled data, synthetic data and unlabeled data without data augmentation. The system that won the 1st place in DCASE 2019 task 4 was the ensemble system of 6 systems with the same model architecture, and the ensemble method is averaging all the probabilities output by the systems.

\section{Method}
\label{sec:pagelimit}
\subsection{Guided learning for semi-supervised SED}
\label{ssec:subhead}
    We use the guided learning method as our basic model framework. The guided learning method is composed of two parts: a professional teacher model (PT-model) and a promising student model (PS-model). 
    The PT-model is designed to achieve reliable audio tagging. As a result, The instance-level feature generated by the PT-model has large receptive field.
    
    The PS-model is designed to detect the sound events, where the audio tags and event boundaries both need to be predicted. Since the PS-model focuses on frame-level pediction, the instance-level feature generated by the PS-model has small receptive field.
    
    During training, the PT-model and PS-model use the same input data. For the data with weak labels in a batch of input data, the PT-model and PS-model both use the labels as their training target. For event category $c$, the loss function is calculated as:
    
    \begin{equation}
    Loss_{\rm{labeled}} = \sum_c cross\_entropy(y_c, \hat{\mathbf{P}}({y_c|\mathbf{x}}))
    \end{equation}
    where $y_c$ is the ground truth.
    
   For the unlabeled data, the PS-model uses the pseudo labels generated by the PT-model as the training target and the PT-model does not have any training target. The pseudo label generated by the
PT-model is obtained as:
    \begin{equation}
        \psi^{\rm{PT}}_c =
        \left\{ \begin{matrix}
        1, &\hat{\mathbf{P}}^{\rm{PT}}(y_c|\mathbf{x}) \geq 0.5 \\
        0, &\rm{otherwise}
        \end{matrix}\right.
        \label{eql}
    \end{equation}
where $\hat{\mathbf{P}}^{\rm{PT}}(y_c|\mathbf{x})$ denotes the probability of audio tagging output by the PT-model. Then the loss function of the unlabeled data is:
    \begin{equation}
    Loss^{\rm{PS}}_{\rm{unlabeled}} = \sum_c cross\_entropy(\psi^{\rm{PT}}_c, \hat{\mathbf{P}}^{\rm{PS}}({y_c|\mathbf{x}}))
    \end{equation}
where $\hat{\mathbf{P}}^{\rm{PS}}({y_c|\mathbf{x}})$ denotes the probability of audio tagging output by the PS-model. 
After $s$ training epochs, the PS-model is able to achieve reliable audio tagging. Then, the audio tagging pseudo labels of unlabeled data output by the PS-model is also used as the training target of the PT-model. The loss function is calculated as:
    \begin{equation}
    Loss^{\rm{PT}}_{\rm{unlabeled}} = \alpha \sum_c cross\_entropy(\psi^{\rm{PS}}_c, \hat{\mathbf{P}}^{\rm{PT}}({y_c|\mathbf{x}}))
    \end{equation}
where $\alpha$ is the hyperparameter to adjust the loss weight. In our experiments, we set $\alpha = 1 - 0.997 ^ {epoch - s}$, $s = 15$.
\subsection{Multi-branch learning for semi-supervised SED}
\label{ssec:subhead}

To further improve the performance, the MBL \cite{MBL} is incorporated into the guided learning system (MBL-GL). Multiple branches with different pooling strategies such as embedding-level pooling and instance-level pooling and different pooling modules such as attention pooling (ATP), global max pooling (GMP) and global average pooling (GAP), are used and share the same feature encoder. As shown in Figure \ref{fig1}, one branch is set as the main branch which takes part in training and detection and another branch is set as the auxiliary branch which is only used for training. 
In our system, we apply the MBL into the PS-model. We choose the embedding-level ATP as the main branch and instance-level GMP or instance-level GAP as the auxiliary branch. The loss function is calculated as:
\begin{equation}
    Loss_{\rm{PS-total}} = a L_{\rm{PS-main}} + b Loss_{\rm{PS-auxiliary}}
\end{equation}
where $a$ and $b$ are the loss weights of the main branch and the auxiliary branch. The $a$ is set to $1.0$ and the $b$ is set to $0.5$ or $1.0$ based on the performance of the validation set.

The reason why we apply the MBL method in the PS-model is that the PS-model outputs the final results of SED while the PT-model only outputs the audio tagging results which is only used in the training process of PS-model. In our early study, we found that the improvement of MBL for audio tagging was limited compared with the improvement for SED.

By using multiple branches, we can also fuse the results of both branches to obtain better result at the inference stage. In this paper, if the auxiliary branch is instance-level GAP, we ensemble the detection results of the main branch and auxiliary branch by taking the average results of instance-level probabilities. 
\begin{equation}
    \hat{\mathbf{P}}_{\rm{fusion}}({y_{ct}}|\mathbf{{x_\textit{t}}}) = 
    \alpha \hat{\mathbf{P}}_{\rm{GAP}}({y_{ct}}|\mathbf{{x_\textit{t}}}) + (1- \alpha)
    \hat{\mathbf{P}}_{\rm{ATP}}({y_{ct}}|\mathbf{{x_\textit{t}}})
\end{equation}
We set $\alpha = 0.5$ in our experiments.

In \cite{MBL}, the MBL approach is proposed only for weakly-labeled data. However, in this work, we need to use the unlabeled data to train our model. In our early experiments, we found that if the ratio between the weakly-labeled data and unlabeled data in each mini-batch was set to be the ratio in the whole training set, which was about 1:9, the MBL-GL performed poorly. The reason for this phenomenon may be that for MBL, although more branches can make the common feature be fit for various learning purposes so that reduce the risk of overfitting, more branches can increase the risk that the common feature can not fit for any learning purpose when the training data contain much noise. For the guided learning framework, the training targets of the unlabeled data for PS-model are produced by the PT-model, which contain noise. To reduce the risk mentioned above, we increase the ratio between the amount of labeled data and unlabeled data to reduce the influence of noise in training data. Besides, different from \cite{MBL}, we only use one auxiliary branch since we find that using two auxiliary branches can decrease the model performance of MBL-GL for the same reason that in guided learning, the training data for the PS-model contain noise and may increase the difficulty to train multiple branches.

\begin{figure}[t]
\centering
\includegraphics[width=0.85\linewidth]{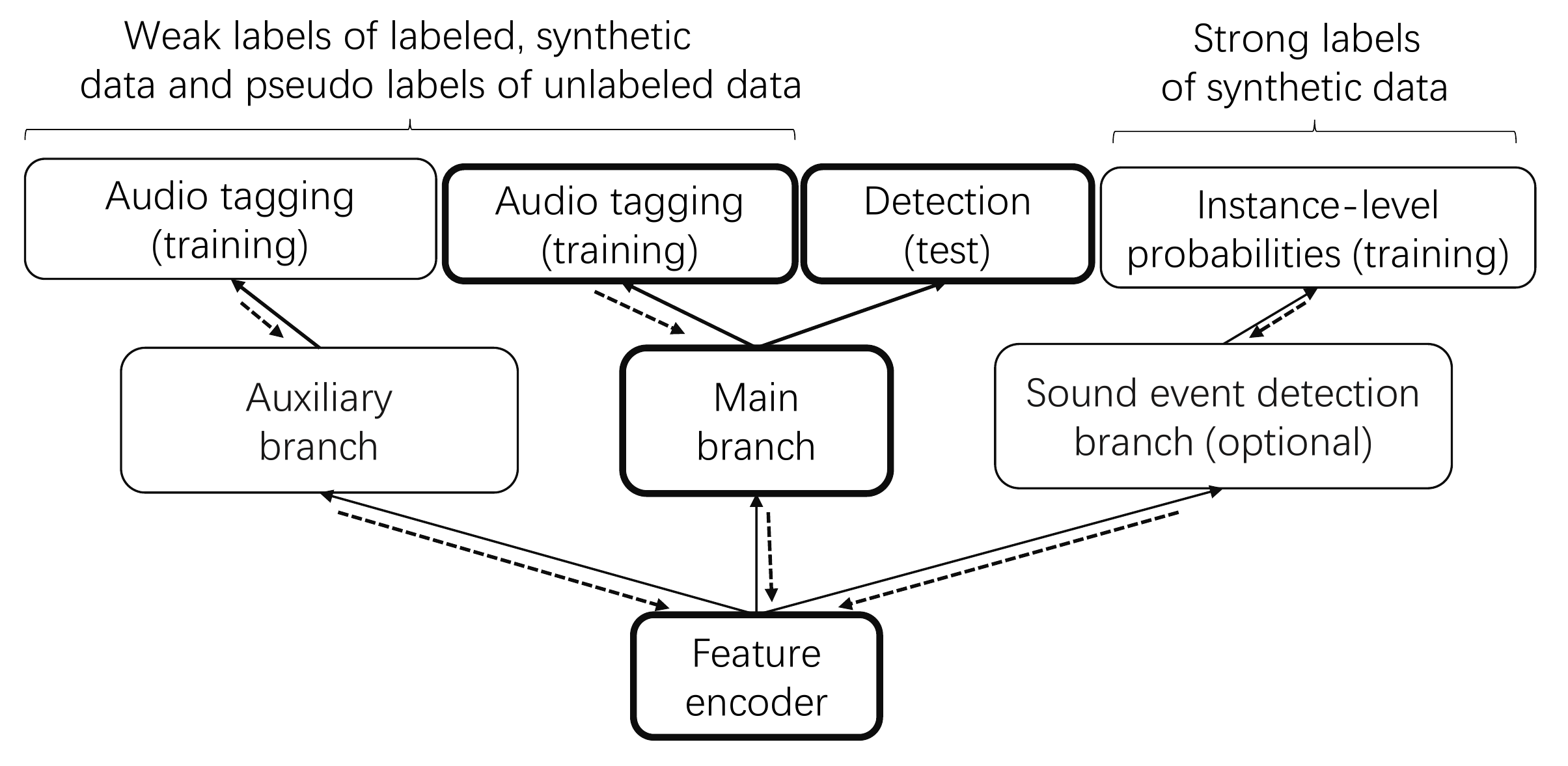}
\vskip -0.12in
\caption{An overview of the architecture of the PS-model}
\label{fig1}
\vskip -0.21in
\end{figure}


\subsection{The detection branch for synthetic data}
\label{ssec:subhead}
In previous study, the multi-task learning of SED in which detecting sound event boundaries and deciding the existence of sound events are considered as two tasks are proved to be a good method to improve the performance of SED. However, this method needs data with strong labels for training. In this work, only the synthetic data has strong labels. To better exploit the synthetic data with strong labels, inspired by multi-task learning, a sound event detection branch (SEDB) is also added. 
As shown in Figure \ref{fig1}, only the synthetic data are used for training the SEDB and the output of the SEDB is the probability of each instance. Then the loss function is calculated as:
\begin{equation}
    Loss_{\rm{SEDB}} = \sum_c\sum_t cross\_entropy(y_{ct}, \hat{\mathbf{P}}({y_{ct}}|\mathbf{x_\textit{t}}))
\end{equation}
where $c$ denotes event category, $t$ denotes frame number,  $\hat{\mathbf{P}}({y_{ct}}|\mathbf{x_\textit{t}})$ is the instance-level probability output by the SEDB and $y_{ct}$ is the instance-level ground truth.
While using the strong labels of the synthetic data to train the SEDB, we also use the weak labels of synthetic data to train other branches of the MBL-GL model.
In our method, we only apply the SEDB in PS-model because the PS-model is mainly used to detect the sound events.

\subsection{Combination of sound separation and SED}
\label{ssec:subhead}

To incorporate sound separation into SED, we train SED models by the separated data output from the baseline system of sound separation. We use the MBL-GL model with instance-level GAP auxiliary branch (no SEDB) as the SS-SED model. Then, we fuse the SED results of models trained by real data and separated data to get the final SS-SED-Ensemble system result (For SS-SED model, the data should be separated and then be used for the SS-SED model at the inference stage). 
\begin{equation}
    \begin{split}
        \hat{\mathbf{P}}^{\rm{SS-SED-Ensemble}}({y_{ct}}|\mathbf{{x_\textit{t}}}) = 
        \sum_i{w_i \hat{\mathbf{P}}_{i}^{\rm{SED}}({y_{ct}}|\mathbf{{x_\textit{t}}})}
        + \\
        \sum_j{w_j \hat{\mathbf{P}}_{j}^{\rm{SS-SED}}({y_{ct}}|\mathbf{{x^{\rm{SS}}_\textit{t}}})}
    \end{split}
\end{equation}
where $\sum_i {w_i} + \sum_j{w_j} = 1$ and $\mathbf{{x^{SS}_\textit{t}}}$ is the feature of separated data.

\section{Experimental Setup}
\label{sec:typestyle}

\subsection{Model architecture}
\label{ssec:Model_architecture}
As shown in Figure 2, for the PS-model, the feature encoder consists of 3 CNN blocks, each of which contains a convolutional layer, a batch normalization layer and a ReLU activation layer. For the PT-model, the feature encoder consists of 9 CNN blocks. The main branch of the PS-model and the PT-model uses embedding-level ATP. And the PS-model has an auxiliary branch which uses instance-level GMP or GAP. The SEDB is optional and is added to the PS-model in some systems. 

\subsection{Data}
\label{ssec:exp_data}
The training set of our system contains
a weakly-labeled set (1578 clips), an unlabeled set (14412 clips), and a strongly labeled synthetic set (2584 clips). The validation set contains 1168 strongly-labeled clips. The public test set contains 692 strong-labeled clips. Data augmentation is also applied in the training process. For all training data, we use time-shifting and frequency-shifting to generate augmented data. For time-shifting, all frames (500) are shifted for exactly 90 steps. For frequency-shifting, all frequencies (64) are shifted for exactly 8 steps. The ratio between original data and augmented data is 8:1.

\subsection{Model training}
\label{ssec:model_training}
We use the Adam optimizer with learning rate 0.0018 to train the model. The learning rate is reduced by 20\% for every 10 epochs. The mini-batch size is set to be 64. For a mini-batch of data, we set the ratio of the weakly-labeled data:synthetic data:unlabeled data to be 3:1:12 (It means that there are 12 weakly-labeled clips, 4 synthetic clips and 48 unlabeled clips in a mini-batch). 

We report the event-based marco F1 score \cite{mesaros2016metrics}. All the experiments are repeated 20 times with random initiation and we report both the average result and the best result of each model.

\subsection{System ensemble}
\label{ssec:subhead}
For system ensemble, we choose different kinds of systems to construct the ensemble system. We take 3 systems with instance-level GAP as auxiliary branch and 3 systems with instance-level GMP as auxiliary branch to construct the SED-Ensemble system. To make the difference between systems large enough, 2 of the 3 systems with instance-level GMP auxiliary branch are with SEDB. To construct the SS-SED-Ensembel system (SS denotes sound separation), besides the 6 systems in the SED-Ensemble system, we add 3 other systems which are trained by sound separated data and has instance-level GAP auxiliary branch.  
We take the weighted sum of all the system outputs as the final results. The ensembling function is: 
\begin{equation}
    \hat{\mathbf{P}}^{\rm{ensemble}}({y_{ct}}|\mathbf{{x_\textit{t}}}) = 
    \sum_i{w_i \hat{\mathbf{P}}_{i}^{\rm{single-system}}({y_{ct}}|\mathbf{{x_\textit{t}}})}
\end{equation}
where $\sum_i{w_i}=1$. The default value of $w_i$ may be $1 / number\_of\_systems$ and in our work, the values are tuned based on the performance on the validation set. Multiple systems with the same model are just trained with different random initializations.

\subsection{The competition systems}
\label{ssec:competition_system}
We participated in 2 subtasks which are SED without SS and SED with SS.
We submitted 4 systems for each subtask and the best system for subtask 1 achieves an event-based F1 score of 44.6\% and the best system for subtask 2 achieves an event-based F1 score of 44.7\%. For our best system of subtask 1, we use 6 systems to make the ensemble system. For our best system of subtask 2, besides the 6 systems in subtask 1, we use 3 systems trained by sound separated data to make the ensemble system.

\section{Experiment Results}
\label{sec:typestyle}

\begin{figure}[t]

\vskip -0.1in
\label{fig2}
\begin{minipage}{0.5\linewidth}
  \centering
  \centerline{\includegraphics[width=0.8\linewidth]{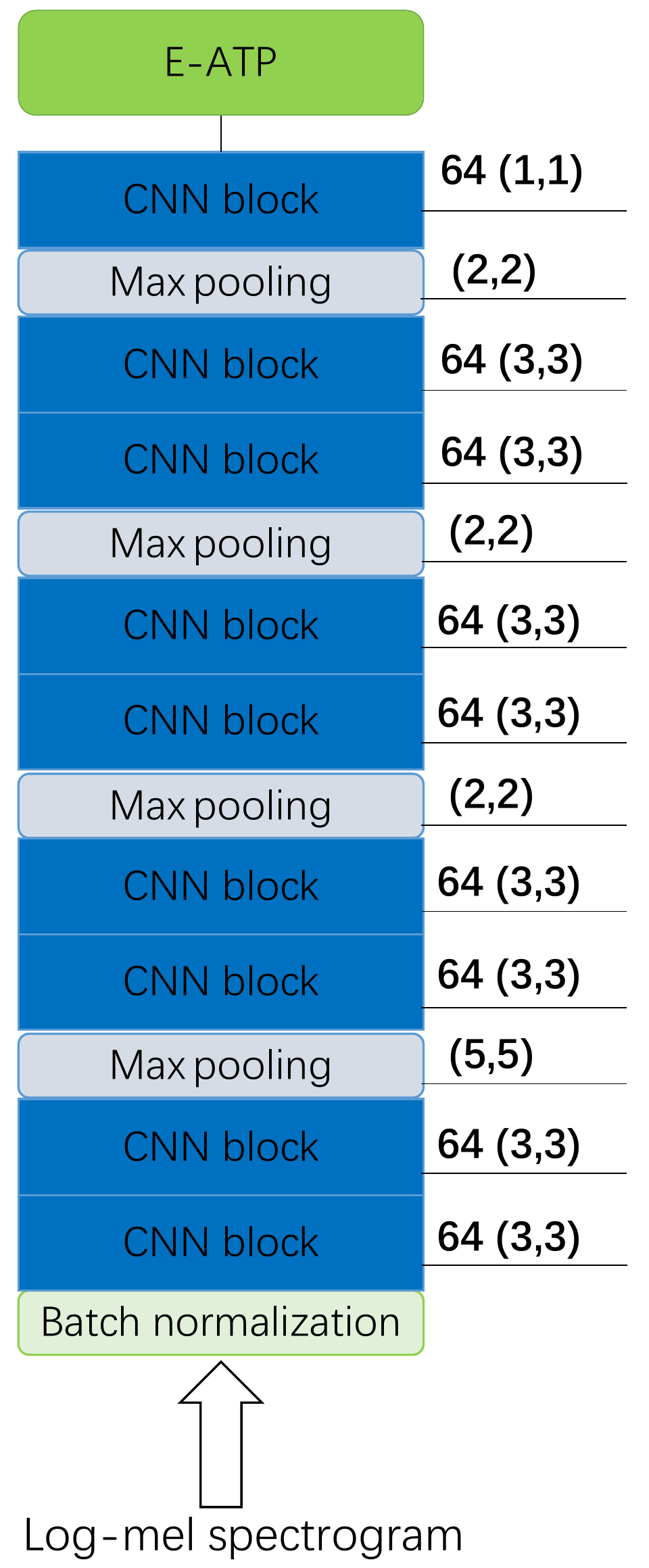}}
  \centerline{(a) PT}\medskip
\end{minipage}
\hfill
\begin{minipage}{0.55\linewidth}
\begin{minipage}{\linewidth}
  \centerline{\includegraphics[width=0.9\linewidth]{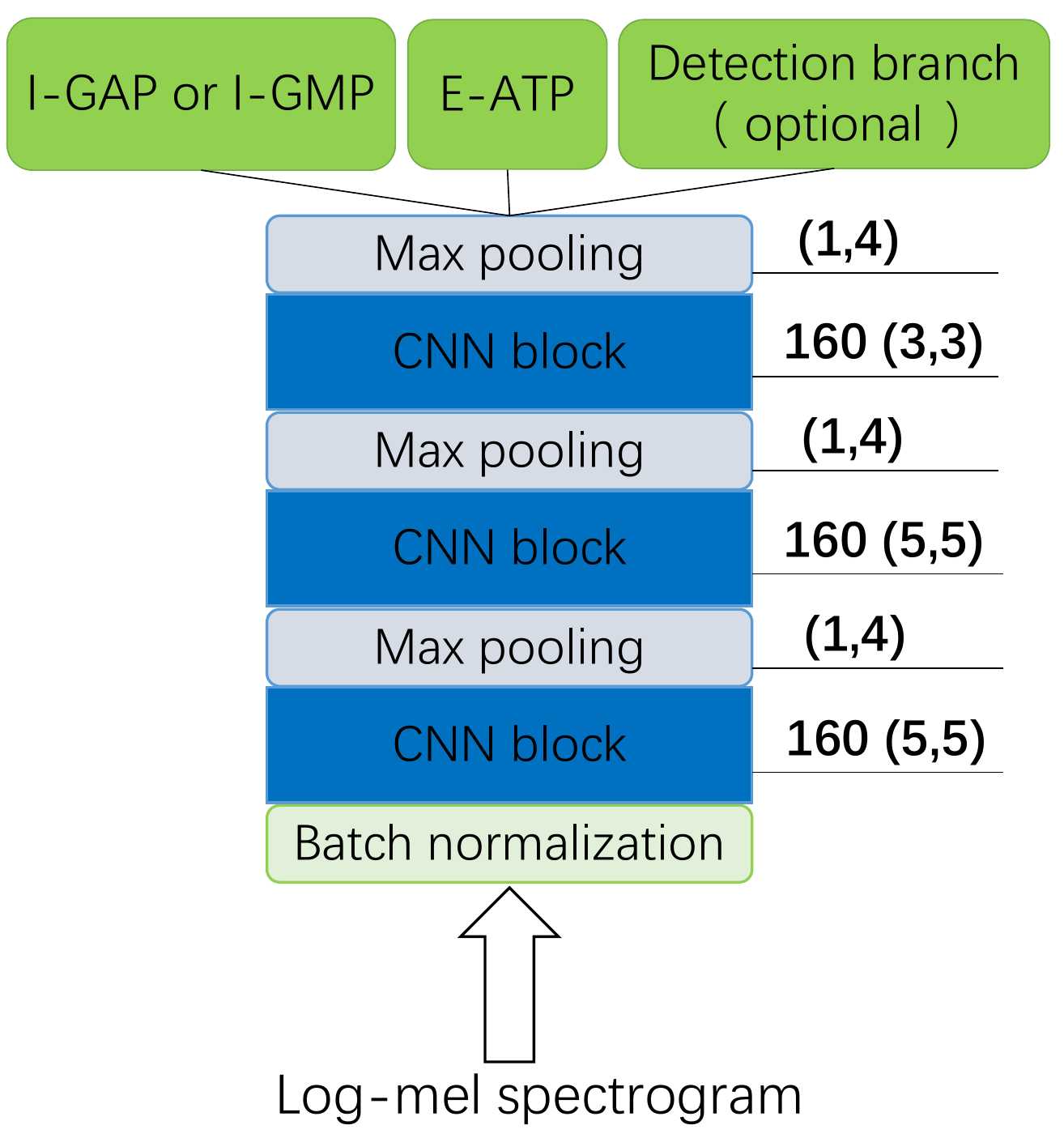}}
  \centerline{(b) PS}\medskip
\end{minipage}

\begin{minipage}{0.9\linewidth}
  \centering
  \centerline{\includegraphics[width=\linewidth]{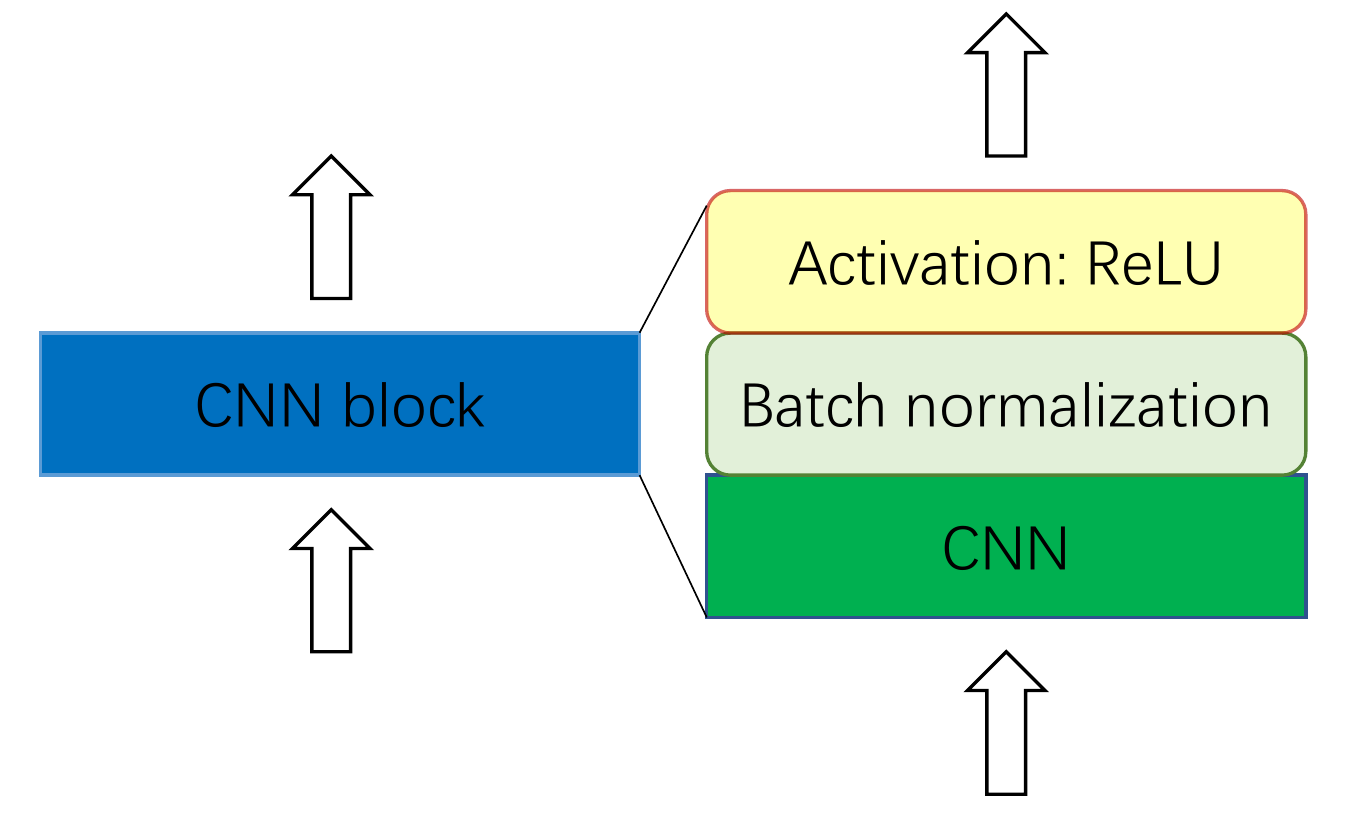}}
  \centerline{(c) CNN Block}\medskip
\end{minipage}
\end{minipage}
 \vskip -0.1in

\caption{The model architectures}
	
\end{figure}

\begin{table}[t]
\vskip -0.25in
  \caption{The event-based F1 of individual systems}
  \vskip 0.1in
  \label{table1}
  \centering
\begin{tabular}{lcc}
\toprule
\textbf{Model}
&\textbf{validation}
&
\textbf{public test}\\
\midrule
\hline
E-ATP + I-GAP-1&$ 0.447 $ & $ 0.463 $\\
E-ATP + I-GAP-2&$ 0.448 $ & $ 0.466 $\\
E-ATP + I-GAP-3&$ 0.451 $ & $ 0.461 $\\
E-ATP + I-GMP-1&$ 0.450 $ & $ 0.466 $\\
E-ATP + I-GMP-2 (with SEDB)&$0.448 $ & $ 0.477 $\\
E-ATP + I-GMP-3 (with SEDB)&$0.454 $ & $ 0.474 $\\
SS - E-ATP + I-GAP-1 & $ 0.378 $ & $ 0.404 $ \\
SS - E-ATP + I-GAP-2 & $ 0.381 $ & $ 0.390 $ \\
SS - E-ATP + I-GAP-3 & $ 0.378 $ & $ 0.394 $ \\
E-ATP + I-GMP-4 (not submitted)&$0.451 $ & $ 0.473 $\\
E-ATP + I-GMP-5 (not submitted)&$0.449 $ & $ 0.474 $\\
E-ATP + I-GAP-4 (not submitted)&$ 0.414 $ & $ 0.441 $\\
E-ATP + I-GAP-5 (not submitted)&$ 0.417 $ & $ 0.439 $\\
E-ATP + I-GAP-6 (not submitted)&$ 0.429 $ & $ 0.439 $\\
\bottomrule 
\end{tabular}
\vskip -0.15in
\end{table}

 Experimental results are shown in Table \ref{table1}, Table \ref{table_valid} and Table \ref{table_test}. Table \ref{table1} shows the results of individual systems which are used for system ensembling, and Table \ref{table_valid} and Table \ref{table_test} show the average and the best results of each kind of system and the results of ensemble systems. For all the experiments, we do not change the PT model and only change the PS models. In the tables, E-* denotes the embedding-level approach and I-* denotes the instance-level approach. SS-* denotes the system uses the sound separated data.
For the baseline system E-ATP, we use the MBL-GL model structure. The only difference between the baseline system and other two kinds of system (E-ATP + I-GAP, E-ATP + I-GMP) is that the baseline system dose not have any auxiliary branch. As shown in Table \ref{table_valid} and Table \ref{table_test}, we find that adding auxiliary branch such as I-GMP or I-GAP can have a beneficial effect. 
For the ensemble system, we use 3 E-ATP + I-GMP and 3 E-ATP + I-GAP systems to construct it. Besides, to make the difference between models larger, 2 of the 3 E-ATP + I-GMP models use SEDB. The ensemble system achieves an F1 score of 0.497 on the public test set and 0.467 on the validation set. Compared to only using systems without the SEDB, using some systems with the SEDB has potential to improve the performance of the ensemble system: We use the 3 E-ATP + I-GAP and 3 E-ATP + I-GMP systems (2 of the 3 systems which use SEDB are replaced by E-ATP + I-GMP-4 and E-ATP + I-GMP-5 ) without SEDB to construct ensemble system which is named SED-Ensemble\_6\_systems. It achieves F1 scores of 0.495 on public test set and 0.463 on the validation set, which are not as good as the ensemble system using SEDB, i. e., SED-Ensemble (submitted).

For the SS-SED-Ensemble system, besides the 6 models used in SED-Ensemble, 3 E-ATP + I-GAP models which are trained by separated data are used and the SS-SED ensemble system achieves F1 scores of 0.495 on the public test set and 0.472 on the validation set. We use 3 other E-ATP+I-GAP systems (E-ATP + I-GAP-4, E-ATP + I-GAP-5, E-ATP + I-GAP-6) trained by real data to replace the SS-SED systems in the SS-SED-Ensemble system. The ensemble system which is named SED-Ensemble\_9\_systems achieves F1 scores of 0.485 on the public test set and 0.463 on validation set, which are lower than the SS-SED-Ensemble system. Although the performances of the 3 E-ATP + I-GAP systems trained by real data are better than the SS-E-ATP + I-GAP systems, they can not improve the system performance while the SS-E-ATP + I-GAP can improve the performance. It proves that adding SS-SED systems to construct the ensemble system can achieve a better performance since the SS-SED systems may have some feature that SED systems do not have. If the performance of the SS-SED system can be further improved, it is expected to further improve the performance of the ensemble system.

\label{ssec:exp_res_valid}
\begin{table}[t]
\vskip -0.08in
  \caption{The event-based F1 scores on the validation set}
  \vskip 0.1in
  \label{table_valid}
  \centering
\begin{tabular}{lcc}
\toprule
\textbf{Model}
&\textbf{Average F1}
&
\textbf{Best F1}\\
\midrule
\hline
E-ATP&$ 0.421\pm 0.0115$ & $ 0.444 $\\
E-ATP + I-GMP&$0.430\pm 0.0088$ & $ 0.445 $\\
E-ATP + I-GAP&$0.431\pm 0.0156$ & $ 0.451 $\\
SED-Ensemble (submitted) & - & $0.467$\\
SED-Ensemble\_6\_systems & - & $ 0.463 $ \\
SS-SED-Ensemble (submitted) & - & $0.472$\\
SED-Ensemble\_9\_systems & - & $ 0.463 $ \\
\bottomrule 
\end{tabular}
\vskip -0.15in
\end{table}

\label{ssec:exp_res_test}
\begin{table}[!t]
\vskip -0.08in
  \caption{The event-based F1 scores on the public test set}
  \vskip 0.1in
  \label{table_test}
  \centering
\begin{tabular}{lcc}
\toprule
\textbf{Model}
&\textbf{Average F1}
&
\textbf{Best F1}\\
\midrule
\hline
E-ATP&$ 0.449\pm 0.0124$ & $ 0.47 $\\
E-ATP + I-GMP&$0.458\pm 0.0125$ & $ 0.478 $\\
E-ATP + I-GAP&$0.450\pm 0.0130$ & $ 0.470 $\\
SED-Ensemble (submitted) & - & $0.497$\\
SED-Ensemble\_6\_systems & - & $ 0.495 $ \\
SS-SED-Ensemble (submitted) & - & $0.495$\\
SED-Ensemble\_9\_systems & - & $ 0.485 $ \\
\bottomrule 
\end{tabular}
\vskip -0.15in
\end{table}

\section{Conclusions}
\label{sec:con}

This paper presents the details of our systems for DCASE 2020 task 4. The systems are based on the first-place system of DCASE 2019 task 4, which adopts the multiple instance learning framework with embedding-level attention pooling and a semi-supervised learning approach called guided learning. The multi-branch learning approach is then incorporated into the system to further improve the performance. Multiple branches with different pooling strategies and different pooling modules are used and share the same feature encoder. To better exploit the synthetic data, inspired by multi-task learning, a sound event detection branch is also added. Therefore, multiple branches pursuing different purposes and focusing on different characteristics of the data can help the feature encoder model the feature space better and avoid over-fitting. The sound separation method is also used and we find that combining the sound separation method to make the ensemble system has great potential to improve the system performance.

\newpage
\newpage
\bibliographystyle{IEEEtran}
\bibliography{refs}

%
%
%
%
%
%
%
%
%

\end{sloppy}
\end{document}